\documentclass[aps,prb,twocolumn,amsmath,amssymb,floatfix,showpacs]{revtex4-1} 
\newlength{\imgwidth}
\usepackage{graphicx} 
\usepackage{psfrag} 
\usepackage{amsmath,amssymb}  
\usepackage{color} 
\usepackage{natbib} 
\newcommand{\I}{\mathrm{i}}
\newcommand{\D}{d}
\newcommand{\E}{\mathrm{e}}

\newcommand{\Br}{{\mathfrak{B}}}
\newcommand{\Tf}{z}
\newcommand{\nl}{n_\mathrm{L}}
\newcommand{\nr}{n_\mathrm{R}}
\newcommand{\ns}{n_\mathrm{S}}

\begin{document} 
\title{Transient noise spectra in resonant tunneling setups:
Exactly solvable models} 
\author{K. Joho} 
\email[To whom correspondence should be addressed: ]{k.joho@thphys.uni-heidelberg.de} 
\author{S. Maier}
\author{A. Komnik} 
\affiliation{Institut f\"ur Theoretische Physik, Universit\"at Heidelberg, Philosophenweg 19, D-69120 Heidelberg, Germany}  
 
\date{\today} 
 
\begin{abstract} 

We investigate the transient evolution of finite-frequency current noise after the abrupt switching on of the tunneling coupling in two paradigmatic, exactly 
solvable models of mesoscopic physics: the
resonant level model and the Majorana resonant level model, which emerges as an effective model for a Kondo quantum dot at the Toulouse point. We find a parameter window in which the transient noise can become negative, a property it shares with the transient current.
However, in contrast to the transient current, which approaches the steady state exponentially fast, we observe an algebraic decay in time of the transient noise for a system at zero temperature. This behavior is dominant for characteristic parameter regimes in both models. At finite temperature the decay is altered
from an algebraic to an exponential one with a damping constant proportional to temperature.
 
\end{abstract} 
%%%%%%%%%%%%%%%%%%%%%%%%%%%%%%%%%%%%%%%%%%%%%%%%%%%%%%%%%%%%%%%%%% 
 
\maketitle

\section{Introduction} 
One central 
issue of mesoscopic physics focuses on 
the transport of charge carriers through nanometer-sized structures where quantum effects 
play an essential role.
In past decades, this research field  experienced a tremendous 
growth.
Not only the electric current but also the  
shot noise, which is associated with 
the charge quantization of current carrying excitations, 
can reveal valuable information about their actual charge.
For instance, the fractional charge of quasiparticles in the fractional quantum Hall regime or the charge of Cooper pairs in superconductors can be recovered in the Fano factor, which is the ratio of the shot noise to electric current.
Nowadays, in addition to the measurement of current-voltage characteristics and noise (current auto-correlation function), 
in many cases even higher cumulants in a non-equilibrium steady state situation can be accessed as well.\cite{basset, ubbelohde} 
The corresponding theoretical tool to gain the information about all current cumulants is referred to as full counting statistics (FCS) and was developed and successfully tested on many free as well as interacting systems during the last 20 years.\cite{lll,Levitov1993,nazarovlong}
However, in preparative non-equilibrium,
where certain parameters are changed rapidly, 
only the current has been extensively addressed so far both experimentally and theoretically.\cite{ensslin,Schmidt2008}
A notable exception has been work [\onlinecite{transient-maci}] on transient current fluctuations at equal times.
At the moment, much effort is 
invested to access the FCS in these situations, but has not been successful even for the simplest models available. 
Instead of following this route, 
we directly calculate the transient finite-frequency noise for two exactly solvable models. 
In particular, we provide a comprehensive analysis of the zero-temperature case
to extract the effects due to shot noise only. 
In addition, the influence of thermal fluctuations 
is addressed for one of the models.
Moreover, our calculations may serve as a benchmark for various numerical simulational methods such as the density-matrix renormalization
group (DMRG), the functional renormalization group (FRG), or  the Monte Carlo technique, which have already been applied 
to some models closely related to those to be treated below.\cite{andergassen,branschaedel,carr,Schmidt2008}

Our paper is structured as follows. In the next section we present our models and the observables of interest. Sections~\ref{SecIII} and~\ref{SecIV} 
are devoted to our main results, namely the analysis of transient noise in the respective models.
Section \ref{SecV} summarizes our findings. The Appendices offer details of several lengthy computations.

\section{Models and Observables}\label{SecII} 
The two models of 
interest are the resonant level model (RLM), which is equivalent to the non-interacting Anderson impurity model (AIM),\cite{anderson}
and the Majorana resonant level model (MRLM), which corresponds to a special parameter constellation 
of the interacting resonant level model (IRLM)
and can be mapped onto the Kondo model at the Toulouse point.\cite{emery-kivelson} These two models are rare examples of exactly solvable systems in non-equilibrium.\cite{caroli,schillerPRB1}
We consider them in a two-terminal setup 
describing a quantum dot that consists of one single electronic level coupled to two electronic reservoirs 
at different chemical potentials. 
In the RLM, the lead electrons are treated as non-interacting fermions (Fermi liquids), whereas in the MRLM, depending on the system realization, they are one-dimensional (1D) interacting fermions (Luttinger liquids) and, in addition, perceive a Coulomb repulsion with an electron on the dot if one starts with a resonant level in a Luttinger liquid,\cite{komnik-ratchet} and non-interacting Fermi liquids in the case of the Toulouse point of the Kondo model.\cite{schillerPRB2} 
A typical realization of the RLM is a quantum dot on the basis of a semiconductor heterostructure in the regime in which electronic correlations on the dot are negligible. Alternatively, one can think of quantum dots in the deep Kondo limit, the transport properties of which are dominated by the resonant level physics.\cite{goldhaber,cronenwett,GlazmanAnkara}
In contrast, the MRLM can be relevant in dot-lead structures composed of single-wall carbon nanotubes,
in which the conduction electrons are strongly correlated and form Luttinger liquids.\cite{bockrath,Egger1997}

\subsection{Resonant level model}
%We follow the notation of (?).
In this case the system under consideration is purely non-interacting, (i.e., charge carriers are non-interacting both in the leads \emph{and}
the dot region). Consequently, the spin degree of freedom is irrelevant and therefore we suppress the corresponding index of fermion operators. Moreover, we assume that the dot region 
can only be occupied by one single electron. 
In real setups, this is justified if the quantum dot is sufficiently small and the spin degeneracy of the energy levels is lifted, 
for instance by applying a strong external magnetic field
so that transport can occur effectively only through one level. 
The Hamiltonian then reads 
\begin{equation}
\hat{H}_{\mathrm{RLM}}=\hat{H}_0+\hat{H}_D+\hat{H}_{T} \, , 
\end{equation}
where $\hat{H}_0$ specifies the contribution of the free lead electrons 
\begin{equation}
\hat{H}_0=\sum_{k,\alpha=L,R}\epsilon_{k,\alpha} c^{\dagger}_{k,\alpha} c_{k,\alpha} \, , 
\end{equation}
with $c_{k,\alpha}$ denoting the annihilation operator of an electron in lead $\alpha$ with momentum $k$. The second ingredient
is the dot Hamiltonian given by
\begin{equation}
\hat{H}_D=\Delta d^{\dagger}d \, .
\end{equation} In addition, we consider tunneling processes between the leads and the dot region,
represented by the corresponding Hamiltonian
\begin{equation}
\hat{H}_{T}=\sum_{\alpha=L,R}\gamma_{\alpha}(t)\left[\psi_{\alpha}^{\dagger}(x=0)d+\mathrm{H.c.}\right] \, ,
\end{equation}
where $d$ and  $\psi_{\alpha}$ are the annihilation operators of the dot and lead electrons, respectively.
We define the operator of the total current through the constriction as 
\begin{equation}
\hat{I}(t)=\frac{\hat{I}_L(t)-\hat{I}_R(t)}{2} \, ,
\end{equation}
where the operator for the current between an individual lead $\alpha$  
and the dot is given by
\begin{equation}
\hat{I}_{\alpha}(t)=\I\gamma_{\alpha}(t)\left[\psi^{\dagger}_{\alpha}(t)d(t)-\mathrm{H.c.}\right].\label{current_op}
\end{equation}
Anticipating the sudden switching of tunneling we consider later, we already included an explicitly time-dependent tunneling amplitude $\gamma(t)$. For reasons of clarity,
we always assume a symmetric coupling $\gamma_L(t)=\gamma_R(t)=\gamma(t)$. The asymmetric case can be treated as well, but the main physical effects 
to be discussed in this paper are unaffected by its concrete choice.

\subsection{Majorana resonant level model}
We now turn to an extension of the RLM that effectively describes interacting systems.
One interesting realization of this model is an interacting resonant level sandwiched between two electrodes in the Luttinger liquid phase at the interaction parameter $g=1/2$.\cite{komnik-resonant-tunnel,komnik-ratchet} In addition it takes into account the Coulomb repulsion between the resonant level and the leads.
The Hamiltonian is given by
\begin{align}
\hat{H}_{\mathrm{IRLM}}=\hat{H}_K+\hat{H}_T+\hat{H}_C \, ,
\end{align}
where 
\begin{align}
\hat{H}_K&=\Delta d^{\dagger}d+\sum_{\alpha=L,R}\hat{H}_0\left[\psi_{\alpha}\right]
\end{align}
is again the kinetic part describing the localized dot level and 1D interacting fermions modeled by the Luttinger liquids, and 
\begin{align}
\hat{H}_T&=\sum_{\alpha=L,R}\gamma_{\alpha}(t)\left[\psi_{\alpha}^{\dagger}(x=0)d+\mathrm{H.c.}\right]
\end{align}
is the usual tunneling part.
The additional term \cite{komnik-resonant-tunnel, komnik-ratchet, komnik-majorana} 
\begin{align}
\hat{H}_C&=\lambda_C d^{\dagger}d\sum_{\alpha=L,R}\psi^{\dagger}_{\alpha}(x=0)\psi_{\alpha}(x=0)
\end{align}
is responsible for the Coulomb repulsion. 
In a general non-equilibrium setting, this model has not been solved exactly so far. However, it has been shown that 
the special choice $\lambda_C=2\pi$ ($=2\pi v_F$ if the Fermi velocity of the lead electrons $v_F\neq 1$) leads 
to a Hamiltonian quadratic in fermionic operators after 
some transformation steps, namely bosonization followed by a unitary transformation and 
re-fermionization.\cite{emery-kivelson} The resulting model, 
which can be mapped onto the Kondo model at the Toulouse point\cite{toulouse}
is called the \textit{Majorana resonant level model (MRLM)} and possesses an exact solution.\cite{schillerPRB2, schillerPRL} After the series of 
transformations mentioned above its Hamiltonian
can be written down in the following way
\begin{align}
\hat{H}_{\mathrm{MRLM}}=\hat{H}_K\left[\xi,\eta,a,b\right]+\hat{H}'_{T}\left[\xi,\eta,a,b\right] \, ,
\end{align}
where 
\begin{multline}
\hat{H}_K\left[\xi,\eta,a,b\right]=\I\Delta ab\\ +\I\int dx\left[\eta(x)\partial_x\eta(x)+\xi(x)\partial_x\xi(x)+V\xi(x)\eta(x)\right]
\end{multline}
governs the dynamics of the free lead Majorana fields $\eta(x)$ and $\xi(x)$ and local dot Majorana fermions $a$ and $b$, 
which are related to the original dot operator by $d=(a+\I b)/\sqrt{2}$,
whereas
\begin{align}
\hat{H}'_{T}\left[\xi,\eta,a,b\right]=-\I\left[\gamma_{+}b\xi(x=0)-\gamma_{-}a\eta(x=0)\right]
\end{align}
is an interaction term modeling couplings between lead and local dot Majorana fermions. Here, we introduced
the coupling constants $\gamma_{\pm}=\gamma_L\pm\gamma_R$.
We take our operator for the total current through the constriction in Majorana fermion representation\cite{komnik-majorana}
\begin{align}
\hat{I}(t)=-\frac{\I}{2}\left[\gamma_+(t)b(t)\eta(t)+\gamma_-(t)a(t)\xi(t)\right] \, ,
\end{align}
with special emphasis on the time dependence of the tunneling coupling.
For the rest of this paper, we also specialize to symmetric coupling in this case and therefore have
$\gamma_{-}(t)=0$ and define $\gamma_{+}(t)=\gamma(t)$. 
It has to be noticed that the splitting of the current into left and right contributions is only reasonable
in our derivation starting from the resonant tunneling setup between Luttinger liquids. In the Kondo picture this is not meaningful since this model describes 
the scattering of conduction electrons off a local impurity, in which tunneling between the electrodes is a single-stage process (electrode-electrode with a spin-flip of the impurity), while electron transmission in the Luttinger set-up is a two-stage process (electrode-dot-electrode). 
One further difference concerns the interpretation of the dot energy in the Kondo case as a local magnetic field.  Thus the dot magnetization in the Kondo picture corresponds to the dot
occupation in the Luttinger setup. 
In most of the calculations presented below only fields at the tunneling point are involved (which means $x=0$), therefore we suppress the spatial coordinate.

\subsection{Noise and current fluctuations}
In contrast to the intuitive nature of current flowing through a conductor, 
one has a certain degree of freedom in defining
the time-dependent current noise spectrum $S(\Omega)$ in a full quantum treatment of a transient problem. 
We use the following, rather general definition which is directly related to the conventional noise definition in the steady state\cite{blanter}
\begin{equation}   \label{dex}
S(\Omega)=\int_\Sigma d(t-t')\E^{\I\Omega(t-t')}S(t,t') \, ,
\end{equation}
with the irreducible current-current correlation function
\begin{equation}
S(t,t')=\left\langle \hat{I}(t)\hat{I}(t') \right\rangle-\left\langle \hat{I}(t) \right\rangle\left\langle \hat{I}(t')\right\rangle \, ,\label{corr}
\end{equation}
which quantifies the fluctuations accompanying the current flow. $\Sigma$ denotes 
the domain in the space of time differences $t-t'$ in which information about the current correlations is available. In the most obvious case of stationary state $\Sigma=(-\infty,\infty)$ and the current correlation function depends on $t-t'$ only. Therefore the noise as defined in Eq.~(\ref{dex}) is time independent. In general $S(\Omega)$ is a time-dependent quantity though, as we shall see later. 
We can express Eq.~(\ref{corr}) in terms of current cross correlators between different
leads $\alpha$ and $\beta$
\begin{equation}
S_{\alpha\beta}(t,t')
=\left\langle \hat{I}_{\alpha}(t)\hat{I}_{\beta}(t')\right\rangle - \left\langle \hat{I}_{\alpha}(t) \right\rangle\left\langle \hat{I}_{\beta}(t')\right\rangle\label{cross-corr}
\end{equation}
so that we obtain the decomposition
\begin{equation}
S(t,t')=\frac{1}{4}\sum_{\alpha,\beta=L,R=\pm}\left(\alpha\beta\right) S_{\alpha\beta}(t,t').
\end{equation}
Throughout this article, we consider the sudden switching on of the tunneling of the form $\gamma(t)=\gamma\theta(t)$, where $\theta(t)$ is the Heaviside step function.
The substitution of new variables $\tau\equiv t-t'$ and $T\equiv t+t'$ effectively restricts the integration range from $-T$ to $T$ and 
finally leads to the \textit{transient} noise formula (emphasizing the explicit time dependence)
\begin{equation}    \label{dex2}
S(\Omega,T)=\int_{-T}^{+T}d\tau \E^{\I\Omega\tau}S(\tau,T) \, ,
\end{equation}
which has to be evaluated for our two cases.
In a steady state, all Green's functions only exhibit a dependence on the time difference $\tau$. Thus, we can immediately carry out 
the $\tau$ integration to access the stationary solution, which has to be equal to the transient noise in the limit of infinite time $T$,  
\begin{equation}
S^{\mathrm{stat}}(\Omega)=\lim_{T\rightarrow\infty}\int_{-T}^{+T}d\tau \E^{\I\Omega\tau}S(\tau,T).
\end{equation}
This relation serves as a consistency check of our results.
The unit of current noise is given by $\pi^2\Gamma G^2_0$, where $G_0=2e^2/h$ is the conductance quantum 
and $\Gamma$ is the hybridization constant expressible in terms of the tunneling amplitude $\gamma$ and the 
electronic density of states of the leads $\rho_0$, which is assumed to be energy independent for the rest of this article. This 
seemingly crude approximation is often called the wide flat band limit. For the RLM, we take the convention 
$\Gamma=2\pi\rho_0\gamma^2$, whereas for the MRLM, we define $\Gamma=\gamma^2/2$ using $\rho_0=1/(2\pi)$ which is required by the 
transformation procedure. 
One particular advantage of the definition 
(\ref{dex2}) 
is that it can be easily applied to the experimental data in the form of time-dependent current traces as presented in Ref.~[\onlinecite{ensslin}]. 
Nonetheless, the solution of the transient problem as shown below can be very efficiently adopted to any other definition of the transient current as well.

\section{Noise in the RLM}\label{SecIII} 
\subsection{Adiabatic noise and transient current evolution}
\begin{figure}[t]
 \includegraphics[width=\imgwidth]{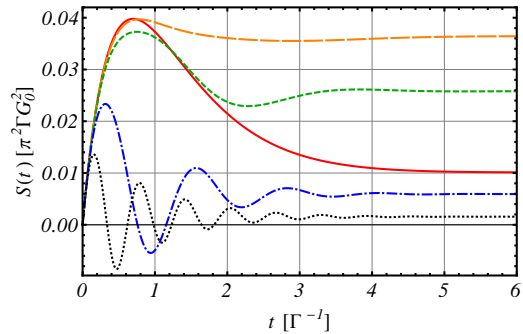}
 \caption{\label{RLM_adiabatic2U}(Color online) Adiabatic noise evolution at fixed voltage $V/\Gamma=1$ and zero frequency $\Omega/\Gamma=0$ for varying 
$\left|\Delta\right|/\Gamma=0, 1, 2, 5, 10$ (red solid, orange long-dashed, green short-dashed, blue dot-dashed, and black dotted curves).}
\end{figure}

Before approaching the problem rigorously, we attempt an approximate calculation of the zero-temperature current noise by assuming that it follows
the transient current adiabatically.  
This \emph{ad hoc} approach can only work well when the corresponding switch-on time $\tau_{\mathrm{sw}}$ is much larger than the typical
time scale of the current evolution, which is proportional to $1/\Gamma$. Nonetheless, we would like to look into the sudden switching case 
$\tau_{\mathrm{sw}}=0$ to obtain a qualitative picture of what might happen to the transient noise.
To achieve our goal, 
we insert the effective \textit{time-dependent} transmission coefficient for the initially empty dot
\begin{widetext}
\begin{align}
\mathcal{T}(\omega,t)=\frac{\Gamma^2-\Gamma \E^{-\Gamma t}(\Gamma\cos\left[(\omega-\Delta)t\right]-(\omega-\Delta)\sin\left[(\omega-\Delta)t\right])}{(\omega-\Delta)^2+\Gamma^2}
\end{align}
\end{widetext}
from the transient current formula given in Ref.~[\onlinecite{Schmidt2008}]
into the generalization of the Schottky formula\cite{schottky} for zero-temperature, zero-frequency (shot) noise in a steady state 
with an energy-dependent transmission coefficient,\cite{khlus,lesovik} which is nothing more than 
the second cumulant of the corresponding charge transfer probability distribution.
Then, we obtain the adiabatic noise evolution as
\begin{align}
S^{\mathrm{adia}}(\Omega=0,t)=\int_{-V/2}^{+V/2}\frac{d\omega}{2\pi}\mathcal{T}(\omega,t)\left[1-\mathcal{T}(\omega,t)\right] \, .
\end{align}
The function $S^{\mathrm{adia}}(t)$ is symmetric with respect to both voltage and dot level energy $\Delta$ and contains only the difference between 
the left and right Fermi functions. 
In general, the time dependence shows up oscillatory behavior with frequencies $V/2\pm\Delta$. On the contrary, the envelope is exponential in time so that
the stationary value is reached after a time proportional to $1/\Gamma$. Of course, this might be just an artifact of our approximation. That is
why we would like to attempt an exact analytical solution of the problem below.
An important point is that for a large enough absolute value of the detuning $\left|\Delta\right|$, the current noise 
according to our definition becomes negative, which is depicted in Fig.~\ref{RLM_adiabatic2U}. This peculiar feature, 
which persists for the transient case to be studied below, has not been reported in the literature so far.
\begin{figure}[b]
 \includegraphics[width=\imgwidth]{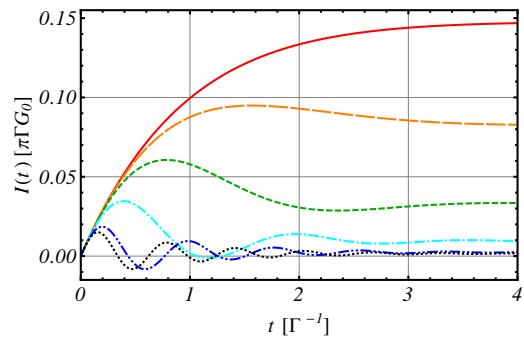}
 \caption{\label{RLM_current_deltaU}(Color online) Transient current at voltage $V/\Gamma=1$ for various detunings $\left|\Delta\right|/\Gamma=0, 1, 2, 4, 8, 10$ 
(red solid, orange long-dashed, green short-dashed, cyan dot-dashed, blue double-dot-dashed, and black dotted curves).}
\end{figure}
We briefly want to turn our attention to 
the total transient current which shares this property, illustrated in Fig.~\ref{RLM_current_deltaU}.
It is due 
to the fact that the transmission coefficient $\mathcal{T}(\omega,t)$, in spite of being properly normalized, 
can become negative.
Although a net charge backflow at intermediate times seems to be counterintuitive at first sight, it can be made plausible since both Fermi levels 
appear to be almost at 
the same height in the case of strong detuning (i.e., when $\Delta$ represents the largest energy scale of all adjustable parameters). 
Of course, this property 
also applies individually to both the left and the right currents. Just after switching on of tunneling the electrons of both leads start to populate the initially empty dot and at the very beginning both $I_R$ \emph{and} $I_L$ have the same sign. Due to the very high energy difference $|\Delta|$ on very short time scales an overpopulation occurs. After that the current signs change and a negative net current can be observed for a rather short time interval. 
\textit{Negative transient current} has already been discussed 
by the authors of Refs.~[\onlinecite{neg_current1}] and [\onlinecite{neg_current2}], 
but in these works, it arises only if the bandwidth of the leads is small enough, whereas in our case, the bandwidth is taken to be infinite. 
Other adiabatic schemes have already been proposed previously (see, for example, Ref.~[\onlinecite{moskalets}] and references therein).
\subsection{Transient noise evolution}
We now want to study
the transient behavior of current noise at finite frequency in its most general form. We compare our results with 
a steady state calculation at finite frequency (corroborated by an FCS calculation at zero frequency).\cite{entin}
In addition, we provide compact formulas for various limiting cases at zero temperature.
The method of choice is the non-equilibrium Keldysh Green's function technique as it provides an intuitive physical picture for every single constituent of relevant equations. As a cross check we performed the same computation using the functional integration technique and obtained precisely the same results. 
The substitution of our current operator Eq.~(\ref{current_op}) into 
Eq.~(\ref{cross-corr}) and assigning times $t$ and $t'$ to different branches of the Keldysh contour, followed by the application of Wick's theorem 
yields\cite{rammer,langreth}
\begin{widetext}
\begin{multline}
S_{\alpha\beta}(t_+,t'_-)=\gamma(t)\gamma(t')%\Theta(t)\Theta(t')
\bigl[G_{dd}^{-+}(t',t)G_{\alpha\beta}^{+-}(t,t')+G_{\beta\alpha}^{-+}(t',t)G_{dd}^{+-}(t,t')\\
-G_{d\alpha}^{-+}(t',t)G_{d\beta}^{+-}(t,t')-G_{\beta d}^{-+}(t',t)G_{\alpha d}^{+-}(t,t')\bigr]
\end{multline}
\end{widetext} 
with the general definition of the Keldysh time-ordered Green's functions
\begin{align}
G_{\zeta\zeta'}(t,t')&=-\I\left\langle T_{C}\psi_{\zeta}(t) \psi_{\zeta'}^{\dagger}(t')\right\rangle\label{coupled}\\
&=-\I\left\langle T_{C}\psi_{\zeta}(t) \psi_{\zeta'}^{\dagger}(t')\hat{S}\right\rangle_0\label{uncoupled}
\end{align}
and the definition of the \textit{S} matrix
\begin{align}
\begin{split}
\hat{S}=T_C \E^{-\I\int_C dt  \hat{H}_T(t)} \, ,
\end{split}
\end{align}
where $\zeta$ and $\zeta'$ specify the respective dot and lead operators. 
Here, we use a compact notation which treats both operators on equal footing.
The average in Eq.~\eqref{coupled} is taken
with respect to the coupled system, while the average in Eq.~\eqref{uncoupled} is performed with respect to the uncoupled one.
The next, somewhat tedious task is to evaluate the various Green's functions. 
To achieve that, we make extensive use of the following general relation for the RLM case, obtained by expansion of the 
\textit{S} matrix to first order and subsequent re-exponentiation,
\begin{widetext}
\begin{align}
\begin{split}
%\begin{multline}
G^{\eta\eta'}_{\alpha\alpha'}(t,t')=g^{\eta\eta'}_{\alpha\alpha'}(t,t')
-\sum_{\sigma=\pm}\sigma\cdot\int_{-\infty}^{+\infty} ds\gamma(s)\left[g^{\eta\sigma}_{\alpha L}(t,s)G^{\sigma\eta'}_{d \alpha'}(s,t')
+g^{\eta\sigma}_{\alpha d}(t,s)G^{\sigma\eta'}_{L \alpha'}(s,t')\right] \, ,
%\end{multline}
\end{split}
\end{align}
\end{widetext}
where the upper indices indicate the branch of the Keldysh contour ($-/+$ for the forward/backward branch) and the lower ones specify the lead/dot operators. 
It proves to be advantageous to express all Green's functions in terms of the \textit{full} dot Green's 
function $D(t,t')\equiv G_{dd}(t,t')$ and the \textit{free} lead Green's functions $g_{\alpha\alpha'}(t,t')$.\footnote{In our terminology,
\textit{free} refers to a dot-lead system without tunneling coupling, while \textit{full} characterizes a coupled system.}
The Dyson equation for the full Keldysh dot Green's function in matrix form is
\begin{multline}
\hat{D}(t,t')=\hat{D}_0(t,t')\\ +\int dt_1\int dt_2 \hat{D}_0(t,t_1)\hat{\sigma}_3\hat{\Sigma}_+(t_1,t_2)\hat{\sigma}_3\hat{D}(t_2,t'),
\end{multline}
where $\hat{\sigma}_3=\mathrm{diag}(1,-1)$ is the third Pauli matrix and, for later use, we defined 
the even/odd tunneling self-energy as
\begin{align}
\begin{split}
\hat{\Sigma}_{\pm}(t,t')=\gamma(t)\gamma(t')\left[\hat{g}_{LL}(t,t')\pm\hat{g}_{RR}(t,t')\right].\label{self-energy}
\end{split}
\end{align}
The free lead Green's function in Fourier-Keldysh space reads\cite{appelbaum, caroli, meir-wingreen1, mahan}
\begin{align}
\hat{g}_{\alpha\alpha'}(\omega)=2\pi\I\rho_0\delta_{\alpha\alpha'}
\begin{pmatrix}
n_{\alpha}-1/2&n_{\alpha}\\
n_{\alpha}-1&n_{\alpha}-1/2
\end{pmatrix},\label{leadgf}
\end{align}
where $n_{\alpha}(\omega)=n_F(\omega-\mu_{\alpha})$ represents the Fermi-Dirac distribution function of the respective lead electrode $\alpha$ with
chemical potential $\mu_{\alpha}$,
while the free dot Green's function is given by 
\begin{widetext}
\begin{align}
\hat{D}_{0}(t,t')=\E^{-\I\Delta (t-t')}
\begin{pmatrix}
-\I[\theta(t-t')(1-n_0)-\theta(t'-t)n_0]& \I n_0\\
-\I (1-n_0)&-\I[\theta(t'-t)(1-n_0)-\theta(t-t')n_0]
\end{pmatrix},
\end{align}
\end{widetext}
where $n_0$ denotes the initial population of the quantum dot.
Using the above relations, we finally obtain the irreducible current-current correlation function
%\begin{widetext}
\begin{align}
S(t,t')=\frac{1}{4}\left[S_{1}(t,t')+S_{2}(t,t')\right],\label{RLMcorr}
\end{align}
where we defined
%\begin{widetext}
\begin{multline}
%\begin{split}
S_{1}(t,t')=D^{-+}(t',t)\Sigma_{+}^{+-}(t,t')\\+\Sigma_{+}^{-+}(t',t) D^{+-}(t,t')
%\end{split}
\end{multline}
and
%\begin{widetext}
\begin{multline}
%\begin{split}
S_{2}(t,t')
=-2\cdot\textrm{Re}\Biggl[\int dt_{1}D^{R}(t',t_{1})\Sigma_{-}^{-+}(t_{1},t)\\ \times \int dt_{2}D^{R}(t,t_{2})\Sigma_{-}^{-+}(t_{2},t')\Biggr].
%\end{split}
\end{multline}
%\end{widetext}
It has to be noted that this formula splits into two major parts. $S_1(t,t')$ involves the \textit{sums} of Fermi functions and depends on the initial dot 
occupation while $S_2(t,t')$ contains the \textit{differences} of Fermi functions and is insensitive to the initial preparation of the system.
To apply this formula, we now have to compute the various dot Green's functions. Thereby, we use the relation $D^{+-}=D^{-+}+D^{R}-D^{A}$ and 
the versions of the Langreth theorem\cite{langreth} for the greater and lesser Green's functions
\begin{align}
\begin{split}
D^{+-}&=(1+D^R\Sigma^R)D^{+-}_0(1+D^A\Sigma^A)+D^{R}\Sigma^{+-}D^{A},\\
D^{-+}&=(1+D^R\Sigma^R)D^{-+}_0(1+D^A\Sigma^A)+D^{R}\Sigma^{-+}D^{A},\label{langreth}
\end{split}
\end{align}
where integration over the internal time variables is implied.
Depending on the initial dot occupation, one of these expressions simplifies tremendously. 
For an initially empty dot $D^{-+}_0(t,t')=0$, whereas for an initially occupied dot we have  $D^{+-}_0(t,t')=0$. 
The retarded and advanced Green's functions were calculated in earlier works\cite{jauho1,jauho2} 
by solving the corresponding Dyson equation, thus we only provide the results for reference
\begin{align}
D^{R}(t,t')&=-\I\theta(t-t')\E^{-\I\Delta(t-t')}e^{-\Gamma(t-t')}
\end{align}
and $D^{A}(t,t')=\left[D^{R}(t',t)\right]^{*}$.
These functions are insensitive to the initial dot occupation, which is reflected in the fact that they are solely dependent on time differences.
In our noise calculations, we first evaluate the time integral to get a formula which explicitly contains the Fermi functions and thus applies at
arbitrary temperatures. We then restrict ourselves to zero temperature 
and give the quite lengthy result in Appendix~\ref{AppA}.
The energy integrals of the first part $S_1(\Omega,T)$ then 
have $-\infty$ as a lower boundary owing to the wide flat band limit, whereas 
the corresponding integrations in the second part $S_2(\Omega,T)$ are performed on compact supports.
The complete finite temperature result is provided in Appendix~\ref{AppB}. 
As an initial condition, we choose an empty dot.

\subsubsection{Steady state solution}
We want to check our results by calculating the steady state noise independently and comparing it later with the limit
$T\rightarrow\infty$ of the transient noise.
Taking advantage of time-translation invariance and transforming to Fourier space, we obtain the following analytical formula
\begin{widetext}
\begin{multline}
S^{\mathrm{stat}}_{\alpha\beta}(\Omega)=\gamma^2\int \frac{d\omega}{2\pi} 
[D^{-+}(\omega)G_{\alpha\beta}^{+-}(\omega+\Omega)
+G_{\beta\alpha}^{-+}(\omega)D^{+-}(\omega+\Omega)\\
-G_{d\alpha}^{-+}(\omega)G_{d\beta}^{+-}(\omega+\Omega)
-G_{\beta d}^{-+}(\omega)G_{\alpha d}^{+-}(\omega+\Omega)].\label{RLMstationary}
\end{multline}
\end{widetext}
Unlike in the time-dependent case, the Green's functions of Eq.~\eqref{RLMstationary} are easily accessible and are obtained by inverting 
the corresponding Dyson equation in matrix form. 
Using another formalism, the steady state noise spectrum for the RLM has first been calculated by the authors of Refs.~[\onlinecite{entin}] and [\onlinecite{orth}], 
which is in excellent agreement with our result. We note in passing
that, in comparison to our graphs, the authors of the aforementioned references obtained mirrored noise spectra with respect to $\Omega$ on account of their 
slightly different definition of the Fourier transformation. As an additional check, we then specialize 
to the case $\Omega=0$, which indeed yields the same stationary result as an independent derivation from the cumulant generating function.

\subsubsection{Limiting cases}
For the zero-temperature shot noise, we give compact, analytical formulas for various limiting cases by holding all other quantities fixed.
The only terms that contribute are those of $S_1(\Omega,T)$. 
For $V\rightarrow\pm\infty$, we obtain
\begin{align}
\lim_{V\rightarrow\pm\infty}S(\Omega,T)=\frac{\Gamma}{4},\label{limit_V}
\end{align}
which is accompanied by the saturation of the total current through the constriction at high voltage,
\begin{align}
\begin{split}
\lim_{V\rightarrow\pm\infty}\left\langle I(t)\right\rangle=\pm\frac{\Gamma}{2}.
\end{split}
\end{align}
These two limits do not display any time dependence. It should be mentioned that this is generally not expected in
a model with finite bandwidth $\epsilon_c$, where the short time scale behavior of the transient current is dominated by oscillations with a period of $1/\epsilon_c$.\cite{Schmidt2008}
For $T\rightarrow0$, we have an offset
\begin{align}
\lim_{T\rightarrow0}S(\Omega,T)=\frac{\Gamma}{4}.\label{RLMoffset}
\end{align}
This limit can be linked to the $V\rightarrow\pm\infty$ case, which is the same, and could thus 
be interpreted as tunneling into vacuum. 
\textcolor{black}{For an arbitrary switching procedure $\gamma(t)=\gamma\theta(t)f(t)$, a detailed 
analysis shows that the offset is generated by a boundary term of the form $\propto f(0)f(T)$, which obviously disappears 
in case of a continuous switching function $f(0)=0$.} 
For $\Omega\rightarrow\pm\infty$, we have
\begin{align}
\lim_{\Omega\rightarrow+\infty}S(\Omega,T)&=\frac{\Gamma}{2},\\
\lim_{\Omega\rightarrow-\infty}S(\Omega,T)&=0.
\end{align}
These are the usual limits of the unsymmetrized noise in the steady state.
We note that the aforementioned cases are all independent of the initial dot occupation.
On the contrary, for $\Delta\rightarrow\pm\infty$, we have for an initially empty dot
\begin{align}
\begin{split}
\lim_{\Delta\rightarrow-\infty}S(\Omega,T)&=\frac{\Gamma}{2}e^{-\Gamma T},\\
\lim_{\Delta\rightarrow+\infty}S(\Omega,T)&=0,
\end{split}
\end{align}
whereas for an initially occupied dot, the limits are reversed. 
This remaining dynamics of noise is understandable 
since, in the former limit, a tunneling process is allowed only for an initially empty dot that can be populated
by one lead electron, whereas in the latter, 
an electron on the dot can jump to one of the leads. This jump probability is equal 
for electrons on/to both leads, thus the time-dependent 
net current vanishes although zero temperature fluctuations are present.
All formulas are clearly in excellent agreement with the steady state result.
\subsubsection{Long-time asymptotics: Zero temperature case}
\begin{figure}[h!]
 \includegraphics[width=\imgwidth]{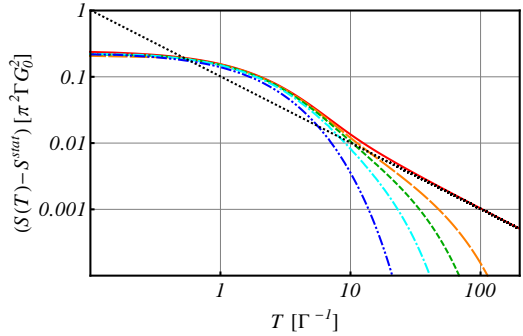}
 \caption{\label{RLM_fit_algebraicU} (Color online) Difference between transient noise and its steady state at $V/\Gamma=\Delta/\Gamma=\Omega/\Gamma=0$ for zero temperature 
(red solid curve) and finite inverse
temperature $\beta\Gamma=200, 100, 50, 20$ (orange long-dashed, green short-dashed, cyan dot-dashed, 
and blue double-dot-dashed curves). We include the reference function $1/(\pi^2 T)$ (black dotted curve).}
\end{figure}

\begin{figure}[htbp!]
 \includegraphics[width=\imgwidth]{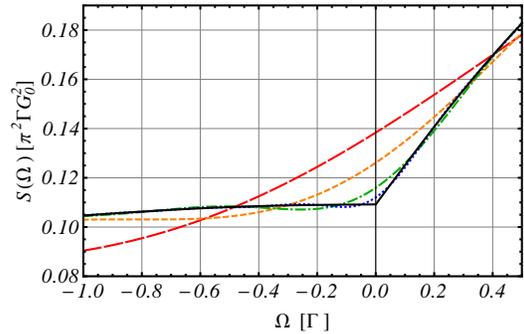}
 \caption{\label{RLM_steady2U} (Color online) Transient noise at times $T\Gamma=2.5, 5, 10, 20$ (red long-dashed, orange short-dashed, green dot-dashed, and blue dotted curves) and steady 
state noise (black solid curve) at $V/\Gamma=2\Delta/\Gamma=10$ as a 
function of frequency $\Omega/\Gamma$.} 
\end{figure}

\begin{figure}[htbp!]
\centering
 \includegraphics[width=\imgwidth]{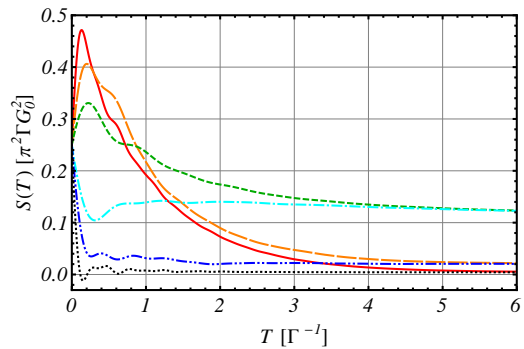}
 \caption{\label{RLM_detuning_pos_neg_corr_colorU} (Color online) Zero-frequency transient noise at fixed voltage $V/\Gamma=10$ for various 
detunings $\Delta/\Gamma=-20, -10, -5, 5, 10, 20$ 
(red solid, orange long-dashed, green short-dashed, cyan dot-dashed, blue double-dot-dashed, and black dotted curves). Note the 
dominance of the algebraic decay of the green and cyan curves ($V=\pm2\Delta$).} 
\end{figure}

\begin{figure}[htbp!]
 \includegraphics[width=\imgwidth]{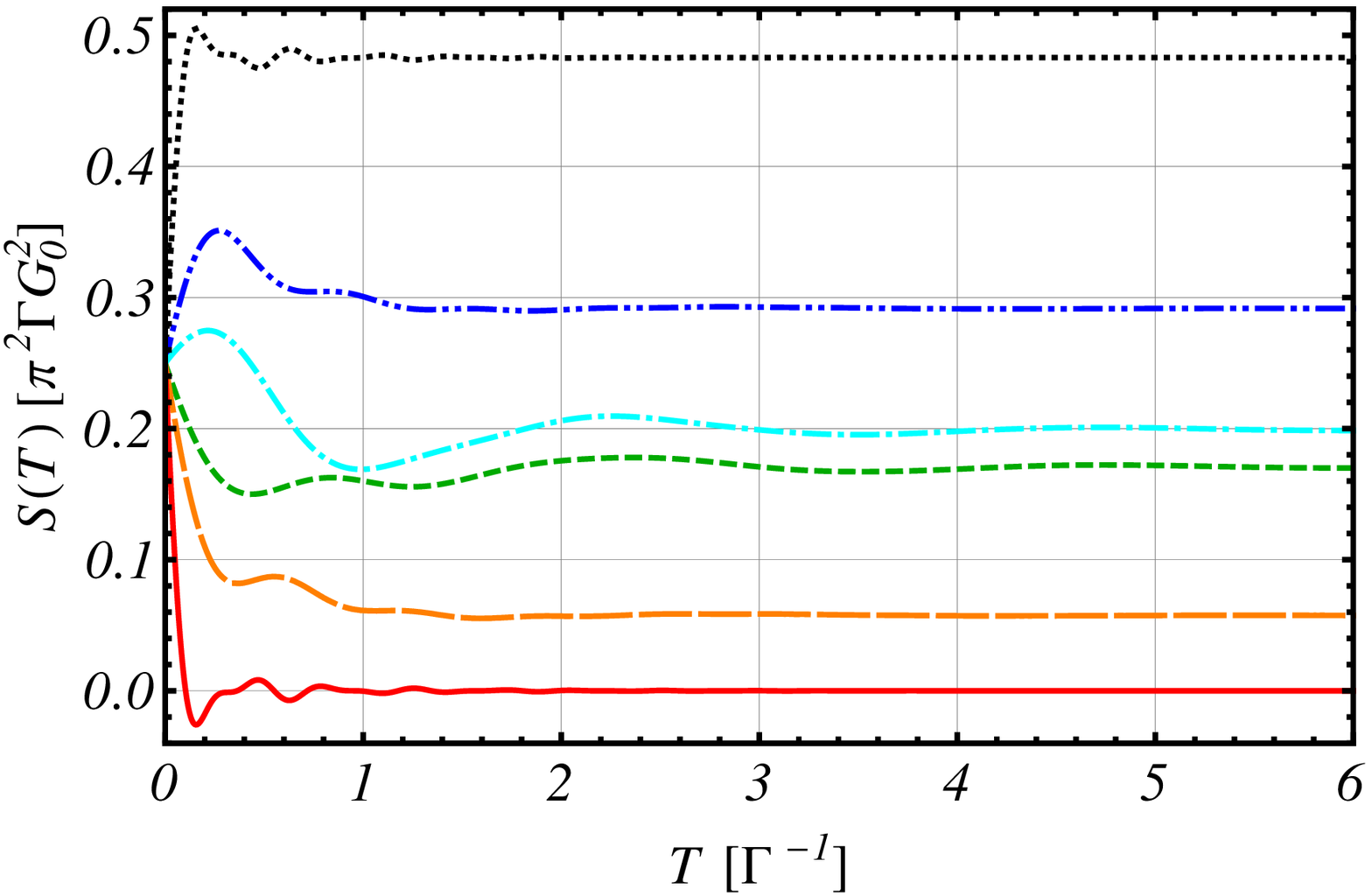}
 \caption{\label{RLM_Omega_pos_neg_V10_colorU}  (Color online) Transient noise at resonance ($\Delta/\Gamma=0$) and fixed voltage $V/\Gamma=10$ for various 
frequencies $\Omega/\Gamma=-20, -5, -2, 2, 5, 20$ 
(red solid, orange long-dashed, green short-dashed, cyan dot-dashed, blue double-dot-dashed, and black dotted curves).}
\end{figure}

\begin{figure}[htbp!]
 \includegraphics[width=\imgwidth]{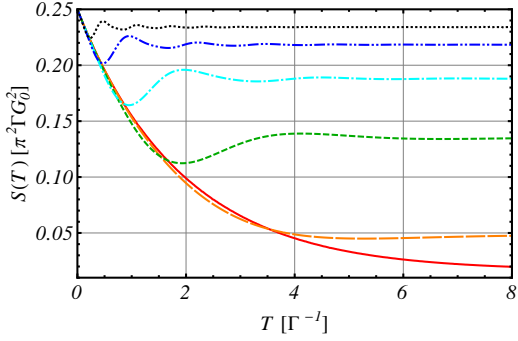}
 \caption{\label{RLM_voltage_final_colorU} (Color online) Zero-frequency transient noise at resonance ($\Delta/\Gamma=0$) for various 
voltages $\left|V\right|/\Gamma=1, 2, 5, 10, 20, 40$ 
(red solid, orange long-dashed, green short-dashed, cyan dot-dashed, blue double-dot-dashed, and black dotted curves).}
\end{figure}

Apart from the special limits above, we now analyze the general long-time behavior of transient noise at zero temperature.
The most astonishing feature is the temporal decay as a power law for large times. 
At zero frequency ($\Omega=0$), we obtain in case of an initially occupied dot
\begin{widetext}
\begin{align}
\begin{split}
S(\Omega=0,T)
=&+\frac{4\Gamma^2}{(4\pi)^2} \sum_{m,n=\pm}\int_{-\infty}^{mV/2} d\omega \frac{\textrm{Si}\left[(\omega+nV/2) T\right]}{(\omega-\Delta)^2+\Gamma^2}
+\frac{\Gamma^2}{4\pi}\sum_{m=\pm}\int_{-\infty}^{mV/2}d\omega\frac{1}{(\omega-\Delta)^2+\Gamma^2}\\
&-\frac{\Gamma^2}{(2\pi)^2}\int_{-V/2}^{V/2} d\omega\int_{-V/2}^{V/2} d\omega'
\frac{\left(\Gamma^2-(\omega-\Delta)(\omega'-\Delta)\right)T\textrm{sinc}\left[(\omega'-\omega)T\right]}{\left[(\omega-\Delta)^2+\Gamma^2\right]\left[(\omega'-\Delta)^2+\Gamma^2\right]}
+g(T),\label{LTA}
\end{split}
\end{align}
\end{widetext}
where Si($x$) is the sine-integral function, sinc($x$) is the cardinal sine function\cite{gradshteyn75} and $g(T)$ comprises all terms which decay 
exponentially and are thus subleading in $T$.
For zero voltage at resonance ($\Delta=0$), this simplifies to produce
\begin{align}
\begin{split}
S(\Omega=0,T)=\frac{\Gamma^2}{2\pi}\int_{-\infty}^{0} d\omega\frac{1+\textrm{Si}\left(\omega T\right)/(2\pi)}{\omega^2+\Gamma^2}+g(T).
\end{split}
\end{align}
To leading order in $1/T$, we find that the transient noise evolution for large times is dominated by a power law
\begin{align}
S(\Omega=0,T\Gamma\gg1)\approx\frac{1}{\pi^2 T}.\label{algeb}
\end{align}
For increasing voltage, this distinctive feature gradually disappears until, at infinite voltage, we attain the limit of Eq.~\eqref{limit_V}. It can only be 
retained 
by adjusting the detuning $\Delta$ in such a way that the Lorentzian peak in the integrand of the first term in Eq.~\eqref{LTA} is shifted 
to the zero of one of the sine integrals (i.e.,~to the position of one of the lead 
Fermi levels $-V/2$ or $V/2$). 
This tendency is illustrated in Fig.~\ref{RLM_fit_algebraicU}. 
Moreover, the feature is only dominant if the frequency fulfills the condition $\Omega/\Gamma\ll 1$, which can be seen in Fig.~\ref{RLM_steady2U} where 
we depict the transient noise spectrum at different times. We note the pronounced discrepancy to the steady state noise spectrum around 
$\Omega/\Gamma=0$, an indicator of the algebraic decay. Apart from that region, the  curves are almost indistinguishable 
for $T\Gamma=20$ on the plotted scale.
In Figs.~\ref{RLM_detuning_pos_neg_corr_colorU}-\ref{RLM_voltage_final_colorU}, we display the effects
of tuning the various parameters of the model, namely voltage, dot level energy and frequency for the case of an initially unoccupied dot.
Obviously, one recognizes the gradual approach to the limits calculated before. 
We stress that, using our definition of noise, we still observe \textit{negative transient noise} in two important cases: large negative frequency or large 
positive/negative detuning for an initially empty/occupied dot, although the steady state noise is always strictly positive. 
This is consistent with very small overall noise levels in the corresponding limiting cases ($\Delta\rightarrow\pm\infty$ and $\Omega\rightarrow-\infty$).
Since at finite values of these parameters, shortly before approaching the extreme cases, we always have oscillatory behavior,
we expect and indeed observe an undershooting below the zero line.
We want to present evidence of a relation between the long-time asymptotics and a feature of the steady state solution.
Indeed, it is striking that the algebraic decay of the transient noise is dominant at zero frequency, where the stationary noise spectrum 
is non-differentiable, its first derivative having a discontinuity $\delta S'$.
\begin{figure}[htbp!]
 \includegraphics[width=\imgwidth]{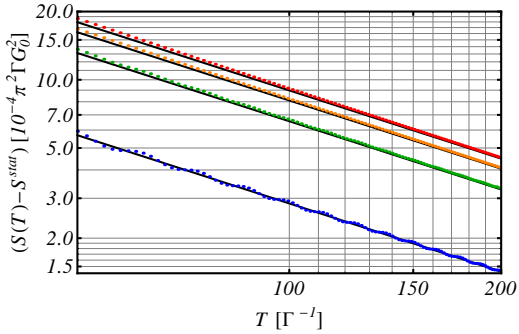}
 \caption{\label{RLM_algeb_steadyU} (Color online) Difference between zero-frequency transient noise and its steady state value 
for parameter pairs $(V/\Gamma,\Delta/\Gamma)=(0.5,0), (0,0.5), (0.5,0.5), (0.5,1)$ (red, orange, green, and blue dotted curves from top to bottom)
compared with the respective reference curves calculated according to the function $\delta S'/(\pi T)$ (black solid curves).}
\end{figure}
Inspired by the plots of Fig.~\ref{RLM_algeb_steadyU}, it is tempting to suggest the following generalization of our 
transient noise formula to arbitrary values of $V$ and $\Delta$,
\begin{align}
S(\Omega=0,T \Gamma \gg 1)=S^{\mathrm{stat}}(\Omega=0)+\frac{\delta S'}{\pi T}+r(T) \, ,\label{disc}
\end{align}
where the discontinuity is given by
\begin{align}
\begin{split}
\delta S'&=\left(\lim_{\Omega\rightarrow 0^+}\frac{\partial S^{\mathrm{stat}}(\Omega)}{\partial\Omega}
-\lim_{\Omega\rightarrow 0^-}\frac{\partial S^{\mathrm{stat}}(\Omega)}{\partial\Omega}\right)\\
&=\frac{1}{2\pi}\sum_{\sigma=\pm}\left(\frac{\Gamma^2}{(\Delta+\sigma V/2)^2+\Gamma^2}\right)^{2-\delta_{V,0}} \, ,
\end{split}
\end{align}
and the function $r(T)$ incorporates all terms of subleading order (i.e., algebraic terms of higher order $\propto 1/T^{\alpha}$ with $\alpha>1$ 
and exponentially decaying functions).
Our conjecture Eq.~\eqref{disc} obviously reproduces our analytical result from Eq.~\eqref{algeb}. Provided it is correct, we conclude the dominance
of the algebraic decay for such parameter constellations in which the dot level coincides with a Fermi level of the electrodes and its gradual disappearance
for growing detuning of the dot level away from a Fermi edge. This is supported by our calculations as well as numerical evaluation, especially by the limiting cases
$V\rightarrow\pm\infty$ and $\Delta\rightarrow\pm\infty$, where the feature is absent.

At this point, we would like to address the similarities and differences to the calculation from Ref.~[\onlinecite{transient-maci}],
which addresses transient equal time current-current fluctuations in an RLM setup.
There the calculated quantity is
\begin{equation}
S(\tau=0,T)=\int_{-\infty}^{+\infty}d\Omega~S(\Omega,T)\label{eqtimecorr} \, ,
\end{equation}
that is, Eq.~\eqref{corr} taken at $t=t'$. Moreover, the $t$ of Ref.~[\onlinecite{transient-maci}] is related to our parameter by $t=T/2$.
The procedure presented there consists of taking a time-dependent bias voltage and assuming its dynamics to be sufficiently slow so 
that an adiabatic approximation can be applied. On the contrary, in our case the 
tunneling coupling is switched on instantaneously and thus infinitely fast and anti-adiabatic.

\subsubsection{Long-time asymptotics: Finite temperature case}
We now want to address the calculation of transient noise for finite temperature. The result obtained after a cumbersome calculation
is provided in Appendix~\ref{AppB}.
We here concentrate on the salient feature which consists of a modification of the temporal decay compared to zero temperature, which is now exponential.
Indeed, we observe that the presence of thermal fluctuations introduces a new energy scale to the problem on which the new damping constant is
linearly dependent. In Fig.~\ref{RLM_fit_algebraicU}, we contrast these two types of decay.  
We point out that in these plots, we have subtracted the respective steady state values due to thermal Johnson-Nyquist noise.
An estimation of the finite-temperature damping constant is provided by
$\Gamma'=\pi/\beta$
so that the envelope of the transient noise for large times is cut off by a function proportional to $\E^{-\pi T/\beta}$, where $\beta$ is the 
inverse temperature. For more details, see Appendix~\ref{AppB}.
This behavior is not unexpected as the transition from algebraic decay at zero temperature to exponential decay at finite temperature
is a quite general phenomenon, which occurs in various systems and is not restricted to temporal evolution. 
As an example, we cite the spatial decay of Friedel oscillations, which follows a similar pattern. 
Furthermore, our result has a dramatic consequence for eventual numerical simulations, which depend sensitively on the approach to steady state. 
We thus conclude that these should be performed at finite temperature to reduce computational effort.
From an experimental point of view, it should be an observable effect, at least at sufficiently low temperature 
where the Fermi functions are not much smeared out so that one can detect the decrease of the damping constant as a function of temperature
in different parameter regimes.
\subsubsection{Correlation function for dot occupation}
We want to mention an interesting similarity to the Fourier transform of the correlation function for the dot occupation,
\begin{equation}
\mathrm{F}(\Omega,T)=\int_{-T}^{+T} d(t-t')\E^{\I\Omega (t-t')}\langle \hat{n}_d(t)\hat{n}_d(t')\rangle.
\end{equation}
In an analogous calculation as before it can be shown that this function already displays an algebraic long-time asymptotics.
\textcolor{black}{For the special case $V=\Delta=\Omega=0$, we find to leading order}
\begin{equation}
\mathrm{F}(\Omega,T\Gamma\gg 1)\approx\frac{2}{\pi^2 T}.
\end{equation}
However, it has to be stated that the charge susceptibility $\chi(\Omega,T)$ exhibits a purely exponential decay in time already at zero
temperature as it is related to a retarded Green's function and thus involves a commutator. Its definition reads 
\begin{equation}
\chi(\Omega,T)=\int_{-T}^{+T} d(t-t')\chi(t,t'),
\end{equation}
where $\chi(t,t')$ is a retarded Green's function given by
\begin{equation}
\chi(t,t')=\I\theta(t-t')\langle \left[\hat{n}_d(t),\hat{n}_d(t')\right]\rangle.
\end{equation}
This behavior is not surprising though. The charge susceptibility represents the response function to external fields. One 
particular realization of such fields is a finite voltage across the constriction. The response is then the current through 
the system which, as we know, has an exponential behavior.

\section{Noise in the MRLM}\label{SecIV} 
We proceed along the lines of the previous Section to evaluate the transient behavior of current noise in the MRLM. 
\subsection{Transient noise evolution}
The transient evolution of the current was calculated in an earlier work.\cite{komnik-majorana} We use the same formalism and define the Majorana Green's function according to the following prescription
\begin{align}
\begin{split}
G_{\zeta\zeta'}(t,t')=-\I\left\langle \zeta(t)\zeta'(t')\right\rangle
=-\I\left\langle \zeta(t)\zeta'(t')\hat{S}\right\rangle_0
\end{split}
\end{align}
with the usual definition of the \textit{S} matrix
\begin{align}
\begin{split}
\hat{S}=T_C e^{-\int_C dt  \gamma(t)b(t)\xi(t)}.
\end{split}
\end{align}
Hence, we obtain the irreducible current-current correlation function
\begin{widetext}
\begin{align}
\begin{split}
S(t_+,t'_-)&=\frac{1}{4}\gamma(t)\gamma(t')\left[G^{+-}_{b\eta}(t,t')G^{+-}_{\eta b}(t,t')-D^{+-}_{bb}(t,t')G^{+-}_{\eta\eta}(t,t')\right]\\
&=\frac{1}{4}\gamma(t)\gamma(t')\left[G^{-+}_{b\eta}(t,t')G^{-+}_{\eta b}(t,t')-D^{+-}_{bb}(t,t')g^{+-}_{\eta\eta}(t,t')\right].
\end{split}
\end{align}
\end{widetext}
In the second line, we used the facts that the retarded mixed Green's function vanishes 
and that the $\eta$-Majoranas decouple from the transport process for symmetric coupling.
For completeness, we write down the retarded Majorana dot Green's function
which is obtained by solving the Dyson equation\cite{komnik-majorana}  
\begin{align}
D^{R}_{bb}(t,t')=-\I\theta(t-t')f(t-t'),
\end{align}
where
\begin{align}
f(t)=\frac{e^{-\Gamma t/2}}{2\Omega'}[(\Omega'-\Gamma/2)e^{\Omega' t}+(\Omega'+\Gamma/2) e^{-\Omega't}],
\end{align}
with $\Omega'=\sqrt{(\Gamma/2)^2-\Delta^2}$.
We also use the Langreth formula Eq.~\eqref{langreth} for $D^{-+}_{bb}(t,t')$ and express the mixed Green's functions 
in terms of the \textit{full} dot Green's 
functions by an expansion of the \textit{S} matrix and subsequent re-exponentiation. 
As a result, we find a similar structure of the irreducible current-current correlation function as in the RLM,
\begin{align}
S(t,t') 
=\frac{1}{4}\left[S_{1}(t,t')+S_{2}(t,t')\right],
\end{align}
where we defined
\begin{align}
\begin{split}
S_{1}(t,t')&=-D^{+-}(t,t')\Xi_{+}^{+-}(t,t')
\end{split}
\end{align}
and 
%\begin{widetext}
%\begin{align}
%\begin{split}
\begin{multline}
S_{2}(t,t')
=\int dt_{1}D^{R}(t,t_{1})\Xi_{-}^{-+}(t_{1},t') \\ \times \int dt_{2}\Xi_{-}^{-+}(t,t_{2})D^{A}(t_{2},t').
\end{multline}
%\end{split}
%\end{align}
%\end{widetext}
The functions $\Xi_{\pm}$ are defined in Fourier-Keldysh space as 
\begin{align}
\hat{\Xi}_{+}(\omega)=\gamma^2\hat{g}_{\eta\eta}(\omega)=\gamma^2\hat{g}_{\xi\xi}(\omega)
\end{align}
and
\begin{align}
\hat{\Xi}_{-}(\omega)=\gamma^2\hat{g}_{\xi\eta}(\omega)=-\gamma^2\hat{g}_{\eta\xi}(\omega)
\end{align}
where the free lead Majorana Green's functions are given by\cite{schillerPRB2,komnik-ratchet}
\begin{align}
\hat{g}_{\eta\eta}(\omega)=\frac{\I}{2}
\begin{pmatrix}
n'_L+n'_R-1&n'_L+n'_R\\
n'_L+n'_R-2&n'_L+n'_R-1
\end{pmatrix}
\end{align}
and
\begin{align}
\hat{g}_{\xi\eta}(\omega)=\frac{1}{2}\left(n'_L-n'_R\right)
\begin{pmatrix}
1&1\\
1&1
\end{pmatrix}.
\end{align}
The primes indicate that, instead of choosing the electrodes' real chemical potentials $\mu_{L,R}=\pm V/2$, we have to insert 
effective ones $\mu'_{L,R}=\pm V$ into the Fermi-Dirac distribution functions.
As expected we recover the stationary state results of the authors of Refs.~[\onlinecite{schillerPRB2}] and  [\onlinecite{komnik-ratchet}].

\subsubsection{Limiting cases}
As for the RLM calculation, we give compact formulas for various limiting cases at zero temperature by holding all other quantities fixed. 
The contributions are due to terms of $S_1(\Omega,T)$, again containing only sums of Fermi functions. 
In the following, we 
list them in the same order as before.
For $V\rightarrow\pm\infty$, we obtain 
\begin{align}
\lim_{V\rightarrow\pm\infty}S(\Omega,T)=\frac{\Gamma}{8}(1+2e^{-\Gamma T}),
\end{align}
which is accompanied by the saturation of the total current through the constriction at high voltage,
\begin{align}
\begin{split}
\lim_{V\rightarrow\pm\infty}\left\langle I(t)\right\rangle=\pm\frac{\Gamma}{4}.
\end{split}
\end{align}
At $T\rightarrow0$, we again have an offset
\begin{align}
\lim_{T\rightarrow0}S(\Omega,T)=\frac{3\Gamma}{8}.
\end{align}
At this point, we would like to mention that the discrepancy of the latter result with Eq.~\eqref{RLMoffset} is due to a 
non-vanishing contribution from the first part
of Eq.~\eqref{langreth} which is absent in the RLM case.
For $\Omega\rightarrow\pm\infty$, we have 
\begin{align}
\lim_{\Omega\rightarrow+\infty}S(\Omega,T)&=\frac{\Gamma}{4}(1+2e^{-\Gamma T}),\\
\lim_{\Omega\rightarrow-\infty}S(\Omega,T)&=0.
\end{align}
However, for $\Delta\rightarrow\pm\infty$, we have for an initially empty dot
\begin{align}
\begin{split}
\lim_{\Delta\rightarrow-\infty}S(\Omega,T)&=\frac{5\Gamma}{8}e^{-\frac{\Gamma T}{2}},\\
\lim_{\Delta\rightarrow+\infty}S(\Omega,T)&=0,
\end{split}
\end{align}
whereas for an initially occupied dot the limits are reversed.
In relation to the RLM case, we state the qualitative difference that we have a temporal
dynamics in the case of limits  $\Omega\rightarrow\pm\infty$ and $V\rightarrow\pm\infty$ for the MRLM. We 
speculate that, at least in the IRLM case, the feature is due to the Coulomb interaction term in the Hamiltonian, which is absent in the RLM case.
The seemingly slower exponential decay in the limits $\Delta\rightarrow\pm\infty$ with $\Gamma/2$ is not directly comparable to the RLM 
due to a different definition of $\Gamma$ in both models. 
Of course, letting $T\rightarrow\infty$ in the above formulas, we find an approach to the expected steady state values.

\subsubsection{Long-time asymptotics: Zero temperature case}
\begin{figure}[htbp!]
 \includegraphics[width=\imgwidth]{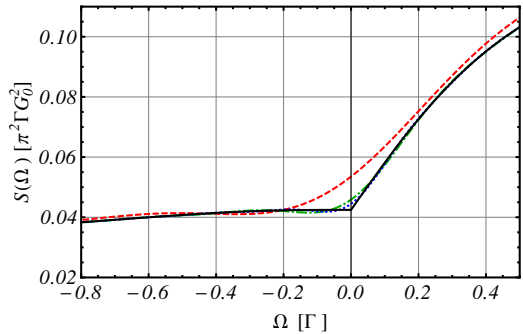}
 \caption{\label{MRLM_steadyU}(Color online) Transient noise at times $T\Gamma=10, 20, 30$ (red dashed, green dot-dashed, and blue dotted curves) and steady 
state noise (black solid curve) at $V/\Gamma=\Delta/\Gamma=5$ as a 
function of frequency $\Omega/\Gamma$.} 
\end{figure}

\begin{figure}[htbp!]
 \includegraphics[width=\imgwidth]{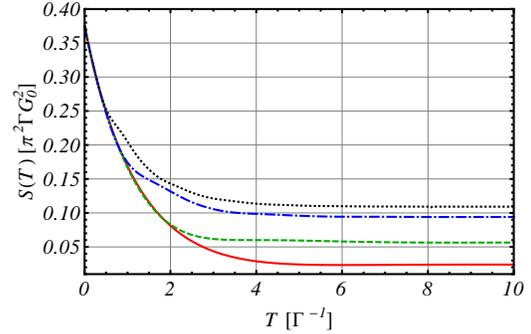}
 \caption{\label{MRLM_voltageU}(Color online) Zero-frequency transient noise at resonance ($\Delta/\Gamma=0$) for various voltages $\left|V\right|/\Gamma=1, 2, 5, 10$ 
(red solid, green dashed, blue dot-dashed, and black dotted curves).}
\end{figure}

\begin{figure}[t]
 \includegraphics[width=\imgwidth]{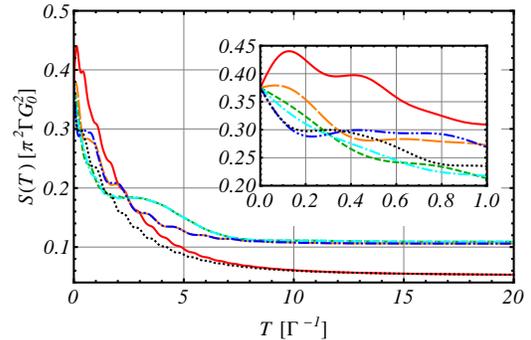}
 \caption{\label{MRLM_detuning_colorU} (Color online) Zero-frequency transient noise at fixed voltage $V/\Gamma=10$ 
for detunings $\Delta/\Gamma=-10, -5, -1, 1, 5, 10$ (red solid, orange long-dashed, green short-dashed, cyan dot-dashed, 
blue double-dot-dashed, and black dotted curves). 
Note the dominance of the algebraic decay of the red and black curves 
($V=\pm\Delta$). The inset shows the same plot zoomed 
in the range $\left[0,1\right]$.} 
\end{figure}

\begin{figure}[t]
 \includegraphics[width=\imgwidth]{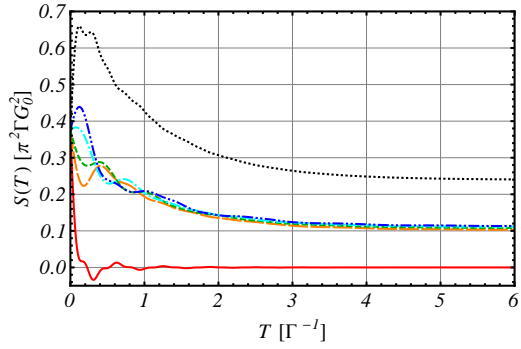}
 \caption{\label{MRLM_Omega_colorU}(Color online) Transient noise at resonance ($\Delta/\Gamma=0$) and fixed voltage $V/\Gamma=10$ for various 
frequencies $\Omega/\Gamma=-20, -5,-2, 2, 5, 20$ (red solid, orange long-dashed, green short-dashed, cyan dot-dashed, 
blue double-dot-dashed, and black dotted curves).}
\end{figure}

In analogy to the RLM case, we 
identify a term with a similar structure involving sine integrals. For $\Omega=0$, it is given by
\begin{widetext}
\begin{align}
\begin{split}
&S_1(\Omega=0,T)\\
&=-\frac{\Gamma^2}{64\Omega'^2\pi^2}\sum_{\sigma=\pm}\int_{-V}^{+V}d\omega\int_{-V}^{+V}d\omega'\sum_{k,m=\pm}\frac{(k\Omega'+\Gamma/2)(m\Omega'+\Gamma/2)}
{\left[\Omega'+k\left(\Gamma/2-\I\omega\right)\right]
\left[\Omega'+m\left(\Gamma/2+\I\omega'\right)\right]}2T\mathrm{sinc}\left[(\omega+\omega')T\right]\\
&\quad+\sum_{\sigma=\pm}\int_{-\infty}^{\sigma V}d\omega\left[\frac{\Gamma^2}{8\pi^2}\sum_{m=\pm}
\frac{\omega^2\left(\textrm{Si}\left[(\omega-mV) T\right]-\frac{\pi}{2}\right)}
{(\omega^2+\Delta^2)^2+\omega^2(\Gamma^2-4\Delta^2)}-\frac{\Gamma}{16\Omega'\pi}\sum_{m,n=\pm}\frac{(\Omega'-n\Gamma/2)}{\I m (\Omega+\omega)+(n\Omega'-\Gamma/2)}\right]+h(T),\label{MLTA}
\end{split}
\end{align}
\end{widetext}
where the function $h(T)$ summarizes all terms that 
are exponentially decaying and thus subleading in $T$.
Note that here, the voltage is doubled with respect to the RLM, a peculiarity due to the transformation steps from the original models.
For $V=\Delta=0$, we come to the same conclusion apart from the prefactor and again obtain an algebraic decay, namely to leading order
\begin{equation} 
S(\Omega=0,\Gamma T\gg 1)\approx\frac{1}{2\pi^2 T}. 
\end{equation}
In general, instead of one Lorentzian peak as for the RLM, the second term
of Eq.~\eqref{MLTA} shows a two-peak structure 
with maxima at $\omega=\pm\Delta$. However, this does not modify our conclusion. Obviously, the term has an appreciable 
effect only if $V\approx\pm\Delta$ (i.e., if the dot level almost coincides 
with one of the~\textquotedblleft dressed\textquotedblright~lead Fermi levels). If the dot and Fermi levels move away from each other, the two peaks are no longer situated at the 
respective zeros of the sine integrals.
For increasing $\Omega$, we also observe the gradual disappearance of this distinctive feature as shown in Fig.~\ref{MRLM_steadyU}.
Moreover, we emphasize that the transient noise as well as the current can also become negative in the MRLM. 
The transient noise evolution for various parameters in the case of
an initially empty dot is shown in Figs.~\ref{MRLM_voltageU}-\ref{MRLM_Omega_colorU}.

\section{Discussion and Outlook}    \label{SecV}
The most striking  
feature that distinguishes the zero temperature transient noise from the evolution of current and dot population in both the RLM and the MRLM
is its algebraic temporal decay dominant for certain parameter sets. It achieves its maximum 
magnitude 
if one of the (dressed in the case of resonant tunneling between Luttinger liquids) Fermi levels matches the dot energy at $\Omega=0$ and is suppressed if one of the model parameters becomes significantly larger than $\Gamma$.
With increasing frequency $\Omega$, the feature also becomes less pronounced.
In both cases of conventional as well as Majorana RLM, this remarkable feature can be traced back to contributions involving energy integrals over sinc functions, which, in turn, are a result of resonances in involved Green's functions.

We expect this effect to survive in the case of realistic band structures beyond the adopted wide flat band limit since a finite bandwidth can only affect the transient behavior on short time scales.
It is also independent of the detailed switching mechanism as
it is an effect at large times. However, we find that finite temperature destroys this effect by introducing a new energy scale
determining the damping constant of the exponential decay. Thus, we expect the results to be observable at sufficiently low, but finite temperature.

The possible avenues for further progress could be a detailed analysis of the impact of the on-dot interactions within the framework of the conventional Anderson impurity model, or a discussion of transient noise in Kondo systems beyond the Toulouse point.

\begin{acknowledgments}
Financial support was provided by the DFG under Grant No. KO 2235/3 and by the ``Enable Fund'', the CQD, and the HGSFP of the University of Heidelberg.
\end{acknowledgments}
\appendix
\section{Analytical expression for transient noise in the RLM: Zero temperature case}\label{AppA}
Below, we give an exact expression for the transient noise in the RLM at zero temperature for an initially unoccupied dot.
The first part involves single integrals on the non-compact supports $[-\infty,\pm V/2]$ and reads
\begin{widetext}
\begin{align}
S_1(\Omega,T)&=\sum_{\sigma=\pm}\int_{-\infty}^{\sigma V/2}d\omega
\left(\frac{s_{1,1}(\omega,\Omega,T)}{(\omega-\Delta+\Omega)^2+\Gamma^2}+\sum_{i=2}^3\frac{s_{1,i}(\omega,\Omega,T)}{(\omega-\Delta)^2+\Gamma^2}
\right)
\end{align}
with
\begin{align}
\begin{split}
s_{1,1}(\omega,\Omega,T)&=\frac{\Gamma}{4\pi}\left(\Gamma-\E^{-\Gamma T}\Gamma \cos\left[(\Delta-\omega-\Omega)T\right]
+e^{-\Gamma T}(\Delta-\omega-\Omega)\sin\left[(\Delta-\omega-\Omega)T\right]\right),\\
s_{1,2}(\omega,\Omega,T)&=\frac{\Gamma^2}{8\pi^2}\sum_{p,q=\pm}q\cdot\left(\textrm{Si}\left[(pV/2+q\omega-\Omega)T\right]
+\textrm{Si}\left[(pV/2+q\Delta-\Omega)T\right]\E^{-\Gamma T}\right),\\
s_{1,3}(\omega,\Omega,T)&=\frac{\Gamma^2}{8\pi^2}\sum_{p,q,s=\pm}q\cdot
\textrm{Si}\left[\left((pV-q(\omega+\Delta)-2\Omega)+\I q s\Gamma\right)T/2\right] \E^{(\I s(\omega-\Delta)
-\Gamma)T/2}.
\end{split}
\end{align}
The second part consisting of double integrals, both on the compact support $[-V/2,+V/2]$, is given by 
\begin{align}
S_2(\Omega,T)&=\sum_{\sigma=\pm}\int_{-V/2}^{V/2} d\omega\int_{-V/2}^{V/2} d\omega'
\sum_{i=1}^4\frac{s_{2,i}(\omega,\omega',\Omega,T)}
{\left[(\omega-\Delta)^2+\Gamma^2\right]\left[(\omega'-\Delta)^2+\Gamma^2\right]}
\end{align}
with
\begin{align}
\begin{split}
s_{2,1}(\omega,\omega',\Omega,T)&=-\frac{\Gamma^2}{4\pi^2}\left[\Gamma^2-(\omega-\Delta)(\omega'-\Delta)\right]
T\textrm{sinc}\left[(\omega'-\omega-\Omega)T\right],\\
s_{2,2}(\omega,\omega',\Omega,T)&=-\frac{\Gamma^2}{4\pi^2}\E^{-\Gamma T}\left[\Gamma^2-(\omega-\Delta)(\omega'-\Delta)\right]
\cos\left[(\omega+\omega'-2\Delta)T/2\right]T\textrm{sinc}\left[(\omega'-\omega-2\Omega)T/2\right],\\
s_{2,3}(\omega,\omega',\Omega,T)&=+\frac{\Gamma^2}{4\pi^2}\E^{-\Gamma T}(\omega+\omega'-2\Delta)\Gamma\sin\left[(\omega+\omega'-2\Delta)T/2\right]
T\textrm{sinc}\left[(\omega'-\omega-2\Omega)T/2\right],\\
s_{2,4}(\omega,\omega',\Omega,T)&=-\frac{\Gamma^2}{4\pi^2}\textrm{Re}\left[\frac{\Gamma^2-(\omega-\Delta)(\omega'-\Delta)-\I\Gamma(\omega-\Delta)(\omega'-\Delta)}
{\I(2\omega-\omega'+\Delta-2\Omega\sigma)+\Gamma}\left(\E^{(\I(\Delta-\omega+\sigma\Omega)-\Gamma)T}-\E^{\I(\omega-\omega'-\sigma\Omega)T}\right)\right].
\end{split}
\end{align}
\end{widetext}
$\,$
%\pagebreak
\section{Analytical expression for transient noise in the RLM: Finite temperature case}\label{AppB}
In the case of finite temperature, everything is much more involved. 
We begin with the analog contribution to $S_1$ in case of zero temperature. 
It splits into two parts, one, where all integrals are performed,
\begin{widetext}
\begin{gather}\begin{split}
&S_{1,1}= \frac{\Gamma}{8\pi}\frac{1}{4\beta\Gamma}\sum_\sigma\Biggl\{4\I\pi\left[\Br_1\left(\Tf,x_{1\sigma}\right)-\Br_1\left(\Tf,-\bar{x}_{1\sigma}\right)\right]
+4\I\pi\E^{-2\Gamma T}\left[\Br_1\left(\Tf,-x_{1\sigma}\right)-\Br_1\left(\Tf,\bar{x}_{1\sigma}\right) \right] \\
&+\left(2\I\pi-x_{1\sigma}\right)\left[\Br_0\left(\Tf,-\bar{x}_{1\sigma}\right) - \E^{-2\Gamma T}\Br_0\left(\Tf,-x_{1\sigma}\right) \right]-\left(2\I\pi +\bar{x}_{1\sigma}\right)\left[\Br_0\left(\Tf, x_{1\sigma}\right) - \E^{-2\Gamma T}\Br_0\left(\Tf,\bar{x}_{1\sigma}\right) \right] \\
&+ 4\beta\Gamma\left[\pi -\I\psi\left(\frac{1}{2}+\frac{\I x_{1\sigma}}{4\pi}\right)+\I\psi\left(\frac{1}{2}-\frac{\I\bar{x}_{1\sigma}}{4\pi}\right)\right] \Biggr\},
\end{split}
\end{gather}
\end{widetext}
and a second one where one frequency integration is left over,
\begin{widetext}
\begin{gather}
\begin{split}
&S_{1,2}=-\frac{\Gamma^2}{4}\sum_{\sigma,\tau=\pm}\sigma\int\frac{\D\omega}{\left(2\pi\right)^2}\frac{\nl\left(\omega\right)+\nr\left(\omega\right)}{\left(\omega-\Delta\right)^2+\Gamma^2}\Biggl\{
2\pi\ns\left(\frac{x_{2\sigma\tau}}{2}\right) \\&+\I\Br_0\left(\Tf,-x_{2\sigma\tau}\right)-\I\Br_0\left(\Tf,x_{2\sigma\tau}\right)+\E^{-\Gamma T}\left[2\pi\ns\left(\frac{x_{3\sigma\tau}}{2}\right)+\I\Br_0\left(\Tf,-x_{3\sigma\tau}\right) -\I\Br_0\left(\Tf,x_{3\sigma\tau}\right)\right]\\
&-\E^{-\Gamma T/2}\E^{-\I T\left(\omega-\Delta\right)/2}\left[2\pi\ns\left(\frac{x_{4\sigma\tau}}{2}\right)+\I\Br_0\left(\Tf,-x_{4\sigma\tau}\right) -\I\Br_0\left(\Tf,x_{4\sigma\tau}\right) \right]\\
&-\E^{-\Gamma T/2}\E^{\I T\left(\omega-\Delta\right)/2}\left[2\pi\ns\left(\frac{\bar{x}_{4\sigma\tau}}{2}\right)+\I\Br_0\left(\Tf,-\bar{x}_{4\sigma\tau}\right) -\I\Br_0\left(\Tf,\bar{x}_{4\sigma\tau}\right) \right]
\Biggr\}.
\end{split}
\end{gather}
\end{widetext}
$\Br_i\left(z,a\right) = B_z\left(\frac{1}{2}+\frac{\I a}{4\pi},-i\right)$ with $B$ denoting the incomplete Beta function,\cite{abramowitz+stegun} $x_{1\sigma} = \beta\left(\sigma V-2\left(\Delta+\Omega\right)-2\I\Gamma\right)$, $x_{2\sigma\tau} = \beta\left(\tau V-2\Omega-2\sigma\omega\right)$, $x_{3\sigma\tau} = \beta\left( \tau V-2\Omega-2\sigma\Delta\right)$, $x_{4\sigma\tau}= \beta\left(\tau V -2\Omega-\sigma \omega-\sigma \Delta - \I\sigma \Gamma\right)$, and $z=\E^{-2\pi T/\beta}$.
$\ns\left(x\right) = n_\mathrm{F}\left(-x/\beta\right)$ is the Sigmoid function\cite{abramowitz+stegun}. 
The analog contribution to $S_2$ is given by
\begin{widetext}
\begin{gather}
\begin{split}
&S_2=-\frac{\Gamma^2}{4}\mathrm{Re}\sum_{\sigma,\tau=\pm}\tau\int\frac{\D\omega}{\left(2\pi\right)^2}\frac{\nl\left(\omega\right)-\nr\left(\omega\right)}{\I\left(\omega-\Delta\right)+\Gamma}\Biggl\{\frac{\beta}{y_{1\sigma\tau}-y'_{1,\tau}}\Bigl[\Br_0\left(z, y_{1\sigma\tau}\right)-\Br_0\left(z,-y_{1\sigma\tau}\right) \\ &-\E^{\frac{-\I\left(y_{1\sigma\tau}-y_{\tau}\right)T}{2\beta}}\Br_0\left(z, y_{\tau}\right) +\E^{\frac{\I\left(y_{1\sigma\tau}-y_{\tau}\right)T}{2\beta}}\Br_0\left(z,-y_{\tau}\right)+2\I\pi\ns\left(\frac{y_{1\tau}}{2}\right)-2\I\pi\ns\left(\frac{y_{\tau}}{2}\right)\E^{\frac{\I\left(y_{1\sigma\tau}-y_{\tau}\right)T}{2\beta}} \Bigr]\\
&-\frac{2\beta\E^{\I\sigma\Omega T}}{y_{2\sigma\tau}-y_{\tau}}\Bigl[\Br_0\left(z,y_{\tau}\right)-\E^{\frac{\I T\left(y_{2\sigma\tau}-y_{\tau}\right)}{2\beta}}\left(\Br_0\left(z, y_{2\sigma\tau}\right)+\psi\left(\frac{1}{2}+\frac{\I y_{2\sigma\tau}}{4\pi}\right)-\psi\left(\frac{1}{2}+\frac{\I y_{\tau}}{4\pi}\right) \right) \Bigr]\\
&-\frac{2\beta\E^{\I\sigma\Omega T}}{y_{3\sigma\tau}-y_{\tau}}\Bigl[\E^{\frac{\I\left(y_{3\sigma\tau}-y_{\tau}\right)T}{2\beta}}\left(\Br_0\left(z,y_{3\sigma\tau}\right)+\Br_0\left(z,-y_{3\sigma\tau}\right)-2\I\pi\ns\left(\frac{y_{3\sigma\tau}}{2}\right) \right)\\
&-\Br_0\left(z,y_{\tau}\right)+\E^{\frac{\I\left(y_{3\sigma\tau}-y_{\tau}\right)T}{\beta}}\left(-\Br_0\left(z,- y_{\tau}\right)+2\I\pi\ns\left(\frac{y_{\tau}}{2}\right) \right) \Bigr]\Biggr\}
\end{split}
\end{gather}
\end{widetext}
where $y_{\tau}=\beta\left(\tau V -2\Delta-2\I\Gamma\right)$, $y_{1\sigma\tau} =\beta\left(\tau V-2\omega-2\sigma\Omega\right)$, 
$y_{2\sigma\tau}=\beta\left(\tau V-2\omega-4\sigma\Omega \right)$ and $y_{3\sigma\tau} =\beta\left(\tau V-\omega-\Delta-2\sigma\Omega-\I\Gamma\right)$.
To find the decay law of the noise correlation with time, one has to investigate term by term. 
First, we notice that all the remaining integrals are convergent even without the Beta functions (all the Beta functions 
are at most constant or decaying as a function of $\omega$). This is because of the overall Lorentzian-like prefactors. 
To estimate the asymptotics due to the Beta functions, the following power series representations turn out to be 
extremely useful ($\left\vert z\right\vert<1$),
\begin{gather}
B_z\left(a,b\right) = z^a \sum_{n=0}^\infty \frac{\left(1-b\right)_n}{n!\left(a+n\right)}z^n
\end{gather}
where $(x)_n = \Gamma\left(x+n\right)/\Gamma\left(x\right)$ is the Pochhammer symbol. In our case, we only need $\left(1\right)_n = n!$ in the case of $x=0$ and $\left(2\right)_n = \left(n+1\right)!$ in the case of $x=-1$. We introduce the following notation $z' = \beta\left(\eta_\omega + \I\xi\Gamma\right)$ where $\eta_\omega$ is a real function of $\omega$ and $\xi$ is a real constant. With $g_\omega$ we denote an arbitrary complex function of the variable $\omega$. Then one obtains
\begin{widetext}
\begin{gather}\begin{split}
g_\omega B_z\left(\frac{1}{2}+\frac{\I z'}{4\pi},0\right) &= g_\omega\E^{-\pi T/\beta}\E^{\frac{\xi\Gamma T}{2}}\E^{\frac{-\I\eta_\omega T}{2}} \sum_{n=0}^\infty \frac{\E^{-2\pi n T/\beta}}{1/2 + n - \xi \Gamma\beta/\left(4\pi\right) + \I \eta_\omega\beta/\left(4\pi\right)} \\
&=g_\omega\E^{-\pi T/\beta}\E^{\frac{\xi\Gamma T}{2}}\E^{\frac{-\I\eta_\omega T}{2}} \sum_{n=0}^\infty \frac{1/2+n-\xi\Gamma\beta/\left(4\pi\right)-\I\eta_\omega\beta/\left(4\pi\right)}{\left(1/2 + n - \xi \Gamma\beta/\left(4\pi\right)\right)^2 + \left( \eta_\omega\beta/\left(4\pi\right)\right)^2}\E^{-2\pi n T/\beta}\\
&=g_\omega\E^{-\pi T/\beta}\E^{\frac{\xi\Gamma T}{2}}\E^{\frac{-\I\eta_\omega T}{2}} \sum_{n=0}^\infty \left(a_n+\I b_n\right)\E^{-2\pi n T/\beta}. 
\end{split}
\end{gather}\end{widetext} 
Now, it is crucial that for every $\beta$ there exists a positive integer $N_\beta$ so that the modulus of the real part $a_n$ 
or the imaginary parts $b_n$ is majorized by $1$ for $n\geq N_\beta$. Hence, the real or imaginary part of the whole expression can
be estimated by a combination of finite polynomial $p_{N_\beta}\left(\E^{-2\pi T/\beta}\right)$  and $\E^{-2\pi N_\beta T/\beta}/\left(1-\E^{-2\pi T/\beta}\right)$. 
It is important to notice that the prefactor $\E^{\xi\Gamma T/2}$ does not lead to an exponential increase in any case. 
The function $g_\omega$ always suppresses this tendency. Although our result is valid for arbitrary temperature, we emphasize that the 
zero temperature limit (i.e.,~$\beta\rightarrow\infty$) is far from being trivial.
\section{Analytical expression for transient noise in the MRLM: Zero temperature case}\label{AppC}
We provide the result for $\Delta\neq\Gamma/2$ in the zero temperature case. Here, $\kappa=\pm$ specifies the initially occupied/empty dot.
\begin{widetext}
\begin{align}
\begin{split}
S_1(\Omega,T)=\sum_{i=1}^2 s_{1,i}(\Omega,T)+\sum_{\sigma=\pm}\int_{\sigma V}^{\infty}d\omega s_{1,3}(\omega,\Omega,T)+\sum_{\sigma=\pm}\int_{-\infty}^{\sigma V}d\omega 
\sum_{i=4}^7 s_{1,i}(\omega,\Omega,T)
\end{split}
\end{align}
with 
\begin{align}
\begin{split}
s_{1,1}(\Omega,T)&=\frac{\Gamma^3}{32\Omega'^2\pi}\sum_{m,n=\pm}\frac{(\Omega'+n\Gamma/2)^2 \E^{(-n\Omega'-\Gamma/2)T}}
{(\Omega'+n\Gamma/2)^2+\Delta^2}\left(\frac{\pi}{2}+\mathrm{Si}\left[(\Omega+mV)T\right]\right),\\
s_{1,2}(\Omega,T)&=\frac{\Gamma^3}{32\Omega'^2\pi}\sum_{m,n=\pm}\frac{\Delta^2\E^{-\Gamma T/2}}{-2\Delta^2+2\I\kappa\Delta n\Omega'}
\left(\frac{\pi}{2}+\mathrm{Si}\left[\left((\Omega+mV)+\I n\Omega'\right)T\right]\right),\\
s_{1,3}(\omega,\Omega,T)&=\frac{\Gamma}{16\Omega'\pi}\sum_{m,n=\pm}\frac{(\Omega'-n\Gamma/2)}{\I m (\Omega-\omega)+(n\Omega'-\Gamma/2)}
\left(\E^{\left(\I m (\Omega-\omega)+n\Omega'-\Gamma/2\right)T}-1\right),\\
s_{1,4}(\omega,\Omega,T)&=\frac{\Gamma^2}{32\Omega'^2\pi^2}\sum_{m,n=\pm}\frac{-(\Omega'+n\Gamma/2)^2 \E^{(-n\Omega'-\Gamma/2)T}}
{(\Omega'+n\Gamma/2)^2+\omega^2}\left(\frac{\pi}{2}+\mathrm{Si}\left[(\Omega+mV)T\right]\right),\\
s_{1,5}(\omega,\Omega,T)&=\frac{\Gamma^2}{8\pi^2}\sum_{m=\pm}\frac{\omega^2}
{(\omega^2+\Delta^2)^2+\omega^2(\Gamma^2-4\Delta^2)}\left(\textrm{Si}\left[(\omega-mV-\Omega) T\right]-\frac{\pi}{2}\right),\\
s_{1,6}(\omega,\Omega,T)&=\frac{\Gamma^2}{32\Omega'^2\pi^2}\sum_{k,m,n,p=\pm}\frac{m(\Omega'+n\Gamma/2)(m\Omega'+n\Gamma/2)}
{\left[\Omega'+n\left(\Gamma/2-\I k\omega\right)\right]\left[m\Omega'+n\left(\Gamma/2+\I k\omega\right)\right]}
\E^{\left(\I k\omega-n\Omega'-\Gamma/2\right)T/2}\\
&\quad\times\left(\frac{\pi}{2}+
\mathrm{Si}\left[\left(\left(2\Omega+p V-\omega\right)+\I k\left(n\Omega'+\Gamma/2\right)\right)T/2\right]\right),\\
s_{1,7}(\omega,\Omega,T)&=\frac{\Gamma^2}{32\Omega'^2\pi^2}\sum_{m,n=\pm}\frac{(\Omega'-\Gamma/2)(\Omega'+\Gamma/2)\E^{-\Gamma T/2}}
{\left[\Omega'-n\left(\Gamma/2-\I\omega\right)\right]\left[\Omega'+n\left(\Gamma/2+\I\omega\right)\right]}\left(\frac{\pi}{2}+\mathrm{Si}
\left[\left((\Omega+mV)-\I n\Omega'\right)T\right]\right),
\end{split}
\end{align}
and
\begin{align}
\begin{split}
S_2(\Omega,T)&=-\frac{\Gamma^2}{64\Omega'^2\pi^2}\sum_{\sigma=\pm}\int_{-V}^{+V}d\omega\int_{-V}^{+V}d\omega'\sum_{i=1}^4 s_{2,i}(\omega,\omega',\Omega,T)
\end{split}
\end{align}
with
\begin{align}
\begin{split}
s_{2,1}(\omega,\omega',\Omega,T)&=\sum_{k,m=\pm}\frac{(k\Omega'+\Gamma/2)(m\Omega'+\Gamma/2)}
{\left[\Omega'+k\left(\Gamma/2-\I\omega\right)\right]\left[\Omega'+m\left(\Gamma/2+\I\omega'\right)\right]}2T\mathrm{sinc}\left[(\Omega-\omega-\omega')T\right],\\
s_{2,2}(\omega,\omega',\Omega,T)&=\sum_{k,m,n=\pm}\frac{-m\left(\Omega'+n\Gamma/2\right)\left(m\Omega'+n\Gamma/2\right)}
{\left[\Omega'+n\left(\Gamma/2-\I k\omega\right)\right]\left[m\Omega'+n\left(\Gamma/2+\I k\omega'\right)\right]}
\E^{\left(\I k\omega-n\Omega'-\Gamma/2\right)T/2}\\
&\quad\times 2T\mathrm{sinc}\left[\left(\left(2\Omega-2\omega'-\omega\right)+\I k\left(n\Omega'+\Gamma/2\right)\right)T/2\right],\\
s_{2,3}(\omega,\omega',\Omega,T)&=\sum_{m=\pm}\frac{(\Omega'+m\Gamma/2)^2}
{\left[\Omega'+m\left(\Gamma/2-\I\omega\right)\right]\left[\Omega'+m\left(\Gamma/2+\I\omega'\right)\right]}
\E^{\left(\I(\omega-\omega')-2m\Omega'-\Gamma\right)T/2}\\
&\quad\times 2T\mathrm{sinc}\left[(2\Omega-\omega'-\omega)T/2\right],\\
s_{2,4}(\omega,\omega',\Omega,T)&=\sum_{m=\pm}\frac{-(\Omega'+\Gamma/2)(\Omega'-\Gamma/2)}
{\left[\Omega'+m\left(\Gamma/2-\I\omega\right)\right]\left[\Omega'-m\left(\Gamma/2+\I\omega'\right)\right]}
\E^{\left(\I\left(\omega-\omega'\right)-\Gamma\right)T/2}\\
&\quad\times 2T\mathrm{sinc}\left[\left(2\Omega-\omega-\omega'+\I 2m\Omega'\right)T/2\right].
\end{split}
\end{align}
\end{widetext}

\end{document}